\documentclass[a4paper,12pt]{article} 
\usepackage{graphicx} 
\usepackage{graphics} 
\usepackage{amsmath} 
\usepackage{amsfonts}  
\usepackage{amssymb} 

\usepackage[colorlinks=true, pdfstartview=FitV, linkcolor=blue,  citecolor=blue, urlcolor=blue]{hyperref}  

\def\l{\left}
\def\r{\right}
\def\beq{\begin{equation}}
\def\eeq{\end{equation}}

\begin{document}

%
\title{Statistical deprojection of intervelocities, interdistances and masses in the Isolated Galaxy Pair Catalog}
\author{{\Large Laurent Nottale$^1$ and  Pierre Chamaraux$^2$}\\ \\
$^1$ LUTH, UMR CNRS 8102, Paris Observatory, 92195 Meudon, France\\ 
laurent.nottale@obspm.fr\\
$^2$ GEPI, UMR CNRS 8111, Paris Observatory, 92195 Meudon, France\\ 
pierre.chamaraux@obspm.fr
}
\maketitle

\begin{abstract}
 In order to study the internal dynamics of actual galaxy pairs, we need to derive the probability distribution function (PDF) of true 3D (orbital) intervelocities and interdistances between pair members from their observed (projected) values, and of the pair masses from Kepler's third law. Our Isolated Galaxy Pair Catalog (IGPC) of 13114 pairs \cite{Nottale2018a} is used here for this research.
 The algorithms of statistical deprojection elaborated in \cite{Nottale2018b} are applied  to these observational data. We derive the orbital velocity PDFs for the whole catalog and for several selected subsamples. The interdistance PDF is deprojected and compared to analytical profiles which are expected from semi-theoretical arguments.
 The PDF of deprojected pair orbital velocities is characterized by the existence of a main probability peak around $\approx 150$ km.s$^{-1}$ for all subsamples of the IGPC as well as for the UGC pair catalog \cite{Chamaraux2016}. The interdistance PDFs of both the projected and deprojected data are described at large distances by the same power law with exponent $\approx -2$. The whole distributions, including their cores, are fairly fitted by King profiles. The mass deprojection yields a mass/luminosity ratio for the pairs of $M/L=(30 \pm 5)$ in Solar units.
 The orbital velocity probability peak is observed at the same value, $\approx 150$ km/s, as the main exoplanet velocity peak, which points toward a possible universality of Keplerian structures, whatever the scale. The pair $M/L$ ratio is just 5 times the standard ratio for luminous matter, which does not require the existence of non-baryonic dark matter in these systems.  
 \end{abstract}
 

\section{Introduction}

In a recent paper \cite{Nottale2018b} we have provided new methods for statistical deprojection of the velocity differences and interdistances between the members of galaxy pairs and we have validated them by numerical simulations.  In the present work we perform the deprojection of intervelocity and interdistance probability distribution functions  (PDFs) in real pairs catalogs.

In the purpose of understanding the true 3D dynamics of galaxy pairs using our deprojection methods, we have constructed two new pair catalogs using well defined criteria and improved observational data. The first one is the UGC Galaxy Pair Catalog (UGPC), a catalog of $\approx 1000$ pairs with high accuracy radial velocities \cite{Chamaraux2016} extracted from Nilson's Uppsala Galaxy Catalog (UGC \cite{UGC}), which has the advantage to be complete in apparent diameter. The second one is the Isolated Galaxy Pair Catalog (IGPC), a catalog of more than $13000$ pairs \cite{Nottale2018a} with galaxy members brighter than absolute magnitude $-18.5$. This catalog was built by using the HyperLEDA data base \cite{HyperLEDA,Makarov2014} in order to identify galaxy pairs and extract their parameters from the large surveys, mainly SDSS \cite{SDSSDR12}.

We intend to obtain statistical informations on the true (3D) physical characteristics of those pairs, in particular the orbital velocities of the pair galaxies (3D velocity of one member with respect to the other one) and  their interdistance, the pair masses and M/L ratios through luminosity and Kepler's third law, and possibly their orbital elements, semi-major axes and eccentricities (which will be studied in a subsequent paper).

Recall that we have at our disposal only one component of the velocity difference (along the line-of-sight) and two components of the interdistance (projection on the sky plane). Moreover, contrary to the star-planet or double star Kepler motion, the galaxy pair problem is meeting an additional difficulty, since we have access to only instantaneous data, for one effective point on each orbit (at the time scale of our observations compared to the orbital period).

Therefore we have devised new statistical methods in order to obtain the PDF of those 3D quantities from the projected ones \cite{Nottale2018b}. It has been pointed out, in particular by Faber and Gallagher \cite{Faber1979}, that the previous methods of analysis of pair dynamics, which did not have at their disposal the three dimensional PDFs, were highly unsatisfactory. We shall show in the present paper that our deprojection methods, once they are applied to real galaxy pair catalogs, solve this problem of a reliable derivation of the statistics of pair dynamical parameters.

In section \ref{sec2} we give the results of the deprojection method applied to PDFs of velocity differences between pair members, for several subsamples of the IGPC, involving different isolation criteria, different accuracies of velocity measurements and different deprojection methods. We complete these results by applying the deprojection to the UGPC. The obtained distributions show varying maxima and minima of probability density, but all of them are characterized by the existence of a dominant probability peak around the same orbital velocity of $\approx 150$ km/s. In section \ref{sec3} we perform the statistical deprojection of the distances between members of the galaxy pairs. Then we compare the  obtained PDF to the projected one and we fit them with simple functions (power laws at large distance). This allows an analytical deprojection which is found to be in agreement with the numerical deprojection. The section \ref{mass} is devoted to the derivation of the pair mass PDF from a deprojection of Kepler's third law and to its comparison with the luminosity PDF, allowing us to derive a mean M/L ratio for these pairs.
The section \ref{discussion} is devoted to a discussion of these results, in particular to a comparison between the orbital velocity distribution of pair galaxies and exoplanets, and section \ref{conclusion} to the conclusion.

\section{Deprojection of pair intervelocities}
\label{sec2}

\begin{figure}[!ht]
\begin{center}
\includegraphics[width=12cm]{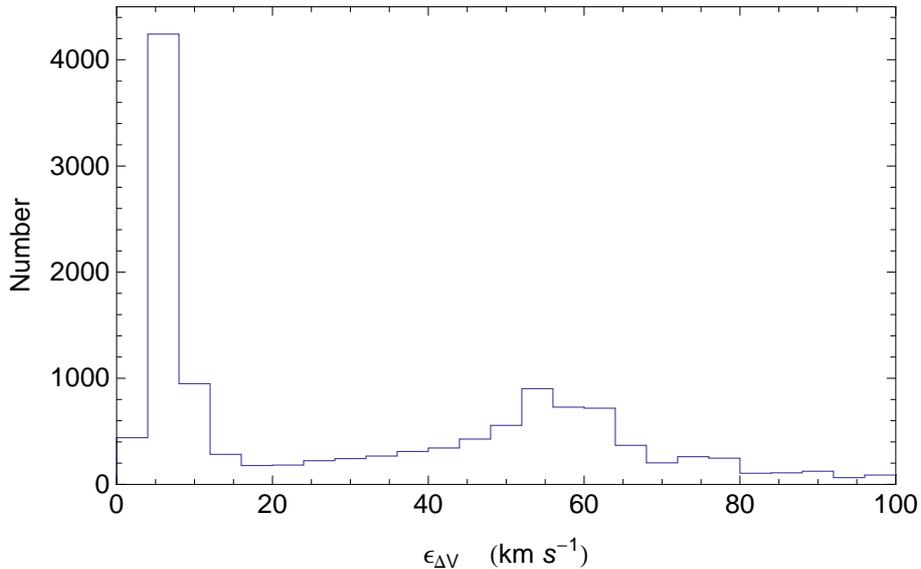}
\caption{{\small  Histogram of the uncertainties on the intervelocities between members of galaxy pairs in the IGP Catalog (13114 pairs). The bin is $5$ km/s. The observed distribution shows two populations, one with accurate velocity errors $<20$ km/s and probability peak $6$ km/s (6026 pairs), and the other with less accurate values having errors peaked around $55$ km/s.}}
\label{Errors}
\end{center}
\end{figure}

We give in Fig.~\ref{Errors} the distribution of uncertainties in the IGPC.  This distribution allows us to define a subsample of 6026 pairs with highly accurate intervelocities ($\delta V<20$ km/s) and a larger subsample of 11259 pairs including less precise values ($\delta V<70$ km/s). We shall use these subsamples in the data analysis in order to check the influence of the velocity accuracy on our results.

The statistical deprojection of intervelocities is performed from the initial PDF of radial velocity differences between pair members. This PDF is given by the histogram of $V=|V_{z2}-V_{z1}|$. Thus the first step in the deprojection consists in the choice of the bin width. Indeed, the $V_z$ PDF has to be strictly monotonous, i.e. decreasing with increasing velocity (with possible plateaux on limited zones, corresponding to hollows in the true (3D) velocity PDF).
 A too small bin width involves fluctuations which may break the expected monotony. Let $N_i$ be the number of points contained in a bin of mean velocity $V_i$ and $N_{i+1}$ in the bin of mean velocity $V_{i+1}$, where $V_{i+1}>V_i$. If $N_{i+1}>N_{i}$,  the computed number of pairs with true intervelocities $(V_i+V_{i+1})/2$ will be negative, which is obviously excluded.
 
\begin{figure}[!ht]
\begin{center}
\includegraphics[width=12cm]{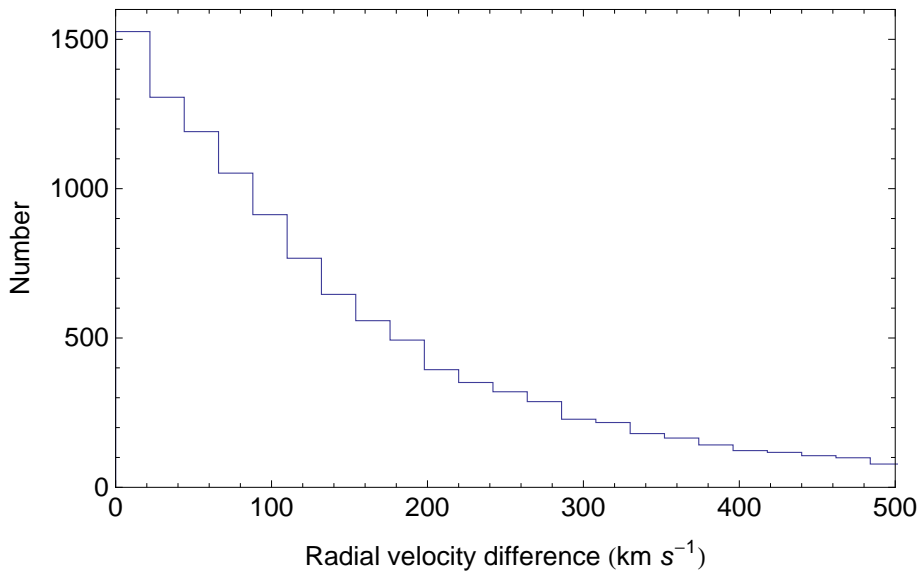}
\caption{{\small  Histogram of the projected intervelocities between members of galaxy pairs in the IGP Catalog, for the subsample of 11259 pairs having errors $<70$ km/s. The bin is $22$ km/s. The observed distribution agrees with the expected monotony of the projected PDF.}}
\label{Vproj}
\end{center}
\end{figure}

 An example of such a monotonous histogram of projected intervelocities for observational data from the IGPC is given in Fig.~\ref{Vproj}. The large number of pairs ($>\approx 10000$ in the IGPC) allows us to use bins of width $22$ km/s, while for smaller catalogs like the UGPC ($\approx 1000$ pairs), the minimal bin width is around $30$ km/s.

 The IGPC has the advantage to provide an information about the degree of isolation of the pairs. We have defined a variable $\rho$ as the ratio $\rho=r_3/r_p$ of the projected distance $r_3$ of the galaxy with $\Delta V<500$ km/s which is closest to the pair center, over the pair member interdistance $r_p$. The catalog contains all pairs with $\rho>2.5$, but it is therefore easy to select subsamples of `fairly isolated pairs' ($\rho>5$, 7449 pairs) or `highly isolated pairs' ($\rho>10$, 4268 pairs), for which specific studies will be carried out.
 
\subsection{Deprojection of highly isolated pair intervelocities}
 The subsample of highly isolated pairs ($\rho>10$) is worth of a specific study, since they can be considered as true two-body Keplerian systems free of external gravitational perturbations.  The result of their statistical deprojection (which provides us with the PDF of their true orbital velocities) is shown in Fig.~\ref{Vdeproj6} for the whole subcatalog and for the subsample of pairs having in addition accurate velocities (mean error $6.4$ km/s).
 
 We use the two-bin difference method and estimate its uncertainty by varying the bin width (left figure of Fig.~\ref{Vdeproj6}).  We also use the moving bin method (right figure) for which we estimate the error by a numerical simulation. This simulation is performed by randomly projecting 50 times the obtained real 3D distribution, then deprojecting again  with the same method each of the projected distributions. This results in the cloud of points shown in Fig.~\ref{Vdeproj6}, which corresponds to a $2\; \sigma$ uncertainty. 
 
Comparing with the fairly isolated sample, we confirm the existence of a main probability peak at $\approx 150$ km/s. A possible secondary peak around 360 km/s is found for the highly isolated pairs, which is not seen in the subsample $\rho>5$.
 
 The PDF shows a fast decrease around 380 km/s, in agreement with the limit we derived from an analysis of the `false' cosmological pair contamination \cite{Chamaraux2016,Nottale2018a}.
 
\begin{figure}[!ht]
\begin{center}
\begin{tabular}{cc}
\includegraphics[width=7.5cm]{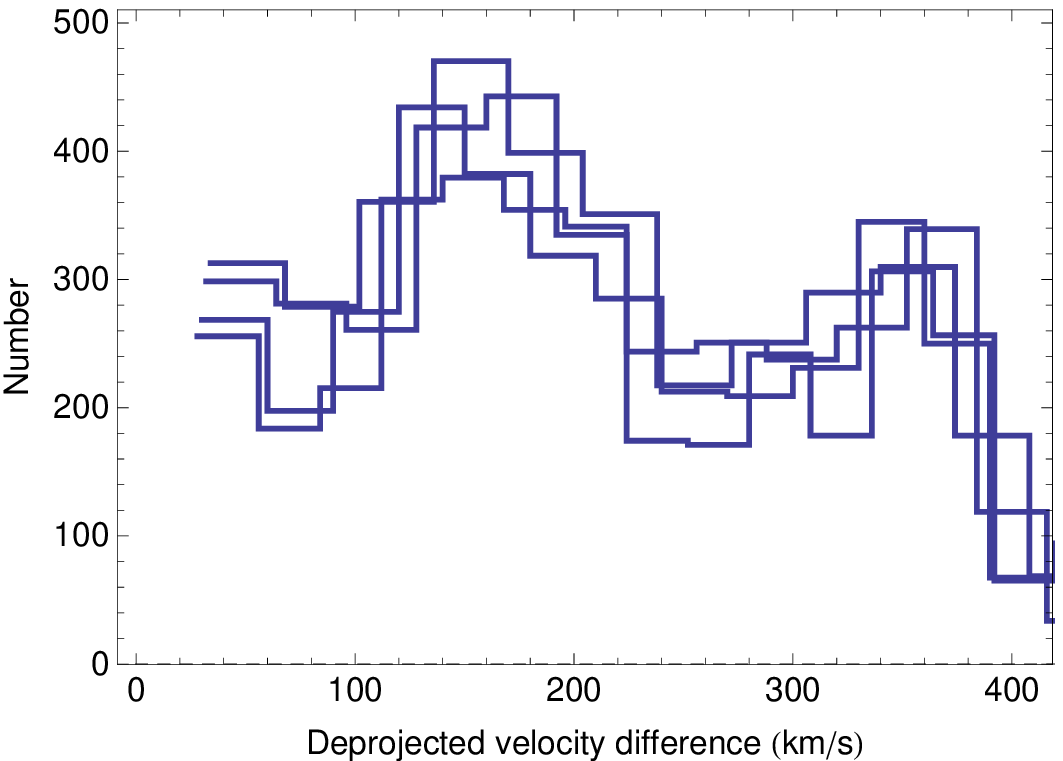} &
\includegraphics[width=7.5cm]{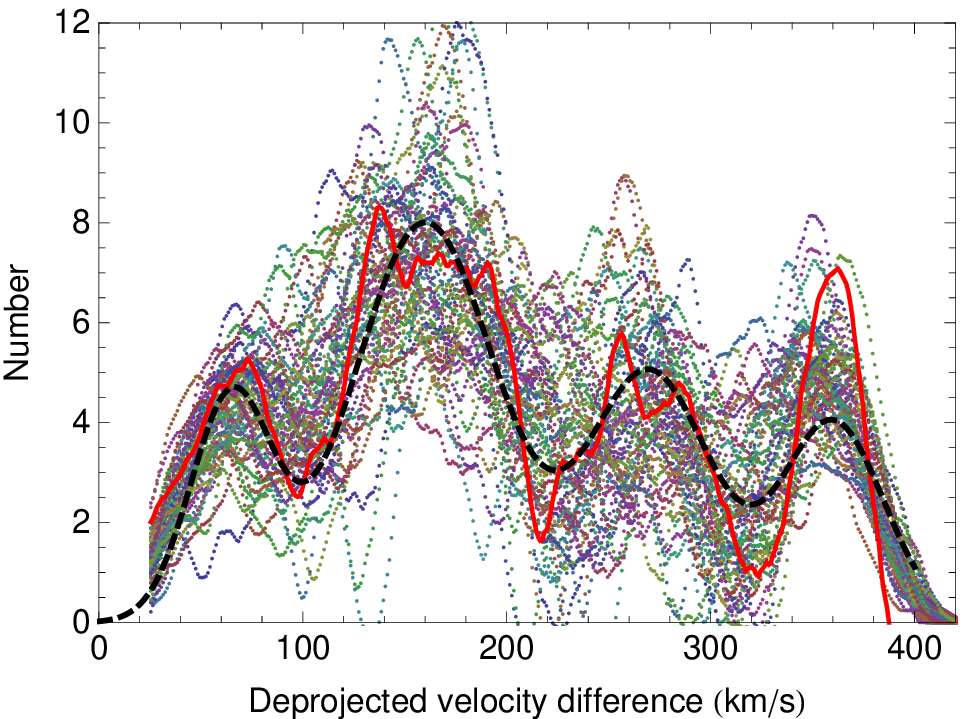}
\end{tabular}
\caption{{\small {\it Left figure:}  PDF of the 3D true intervelocities between members of galaxy pairs (i.e. orbital velocities of the galaxies in the pairs) in the IGP Catalog, deprojected from the radial velocity differences, for the subsample of 4268 highly isolated pairs  $\rho>10$ (see text). The deprojection method used here is the two-bin difference, with bins of $28,\;30,\;32$ and $34$ km/s. Whatever the bin width, probability peaks at $\approx 150$ km/s and $\approx 360$ km/s appear in a stable way. {\it Right figure:} deprojected PDF of the 3D true intervelocities between members of galaxy pairs in the IGP Catalog, for the subsample of highly isolated pairs $\rho>10$, with an additional selection having accurate velocities (errors $<20$ km/s, 1859 pairs). The deprojection uses the moving bin method (the numbers are here the density in effective bins of 1 km/s). The error is estimated from 50 numerical simulations (see text). The two figures agree on the existence of the main peak at 150 km/s and possibly another lower one at $\approx 360$ km/s.}}
\label{Vdeproj5}
\end{center}
\end{figure}

 \subsection{Deprojection of fairly isolated pair intervelocities}
 
\begin{figure}[!ht]
\begin{center}
\begin{tabular}{cc}
\includegraphics[width=7.5cm]{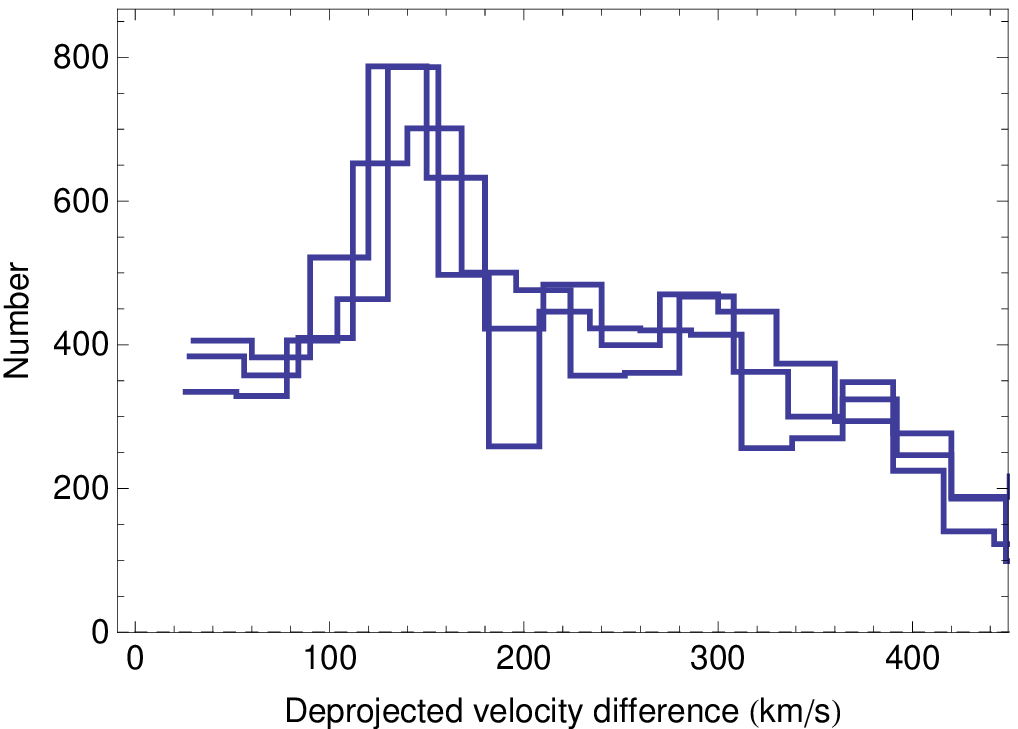} &
\includegraphics[width=7.5cm]{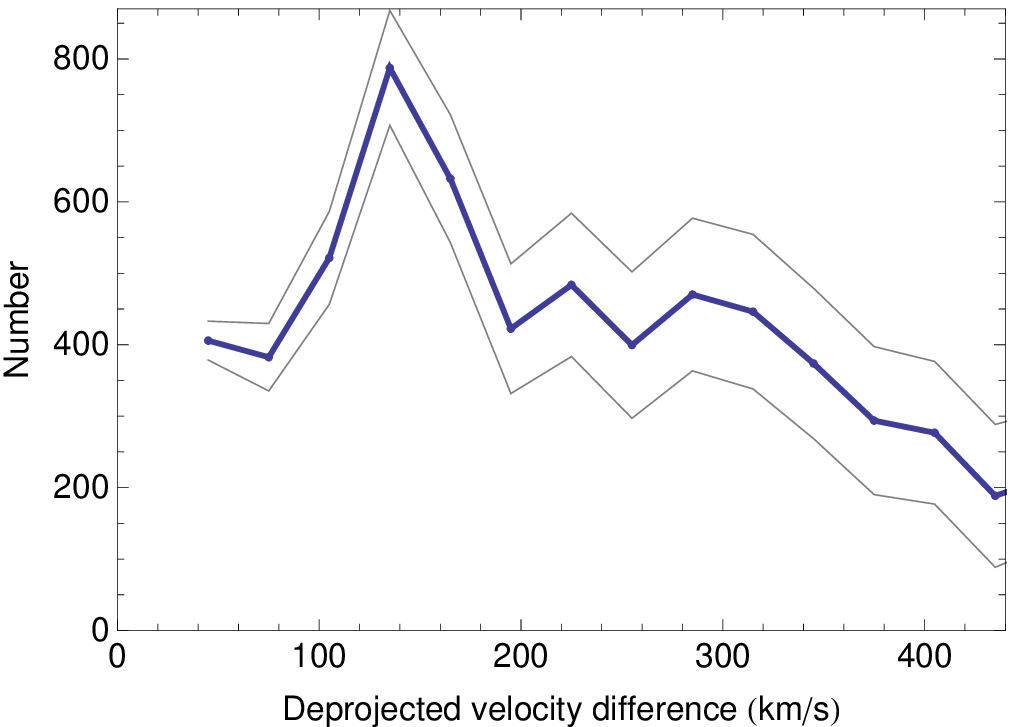}
\end{tabular}
\caption{{\small  PDF of the 3D true intervelocities between members of galaxy pairs in the IGP Catalog (deprojected from radial velocity differences), for the subsample of 7449 fairly isolated pairs $\rho>5$ (see text). {\it Left figure:} the deprojection method used here is the two-bin difference, with bins of $26,\;28$ and $30$ km/s. {\it Right figure:} two-bin difference deprojection method, for a bin width $30$ km/s, accompagnied by the estimated uncertainties  derived from numerical simulations ($\pm 1\, \sigma$, gray curves). The uncertainties derived from the simulation clearly agree with the fluctuations observed for different bins in left figure. Whatever the bin width and the method, a probability peak at $\approx 150$ km/s appears in a stable way.}}
\label{Vdeproj6}
\end{center}
\end{figure}

Let us first apply the two-bin deprojection method to the sub-catalog with $\rho>5$. 

The deprojection formula reads \cite{Nottale2018b}:
\beq
P_v(v)= -v  \l[ \frac{d P_{v_z} (v_z)}{dv_z}  \r]_v,
\label{deproj-v}
\eeq

The simplest way to implement Eq.~(\ref{deproj-v}) consists of :

\noindent (1) constructing the histogram $N^p_i$ of radial (projected) velocities $V_r$ in bins $[V_{i-1},\, V_i]$ of given width $\delta V$;

\noindent (2) computing the differences $(N^p_i-N^p_{i-1})$ between the numbers in successive bins;

\noindent (3) multiplying by the rank $i=V_i/\delta V$ of the bin.

Actually, this method where the difference is taken between two adjacent bins is not optimized and it can therefore be improved. It is more efficient (as in finite difference methods) to take differences between two intervals separated by one bin, $N^p_{i+1}-N^p_{i-1}$. This improvement is based on the fact that $f(x+dx)-f(x)= f'(x) dx + {\cal O}(dx^2)$, while $(f(x+dx)-f(x-dx))/2= f'(x) dx + {\cal O}(dx^3)$.

The uncertainties of this deprojection method can be estimated from numerical simulation. We have found \cite{Nottale2017} that they are half that obtained with the adjacent-bin method), thus achieving a significant improvement.

The result of this deprojection is given in Fig.~\ref{Vdeproj6}.

 \subsection{Deprojection of pair intervelocities (all isolation parameters)}
 
\begin{figure}[!ht]
\begin{center}
\begin{tabular}{cc}
\includegraphics[width=7.5cm]{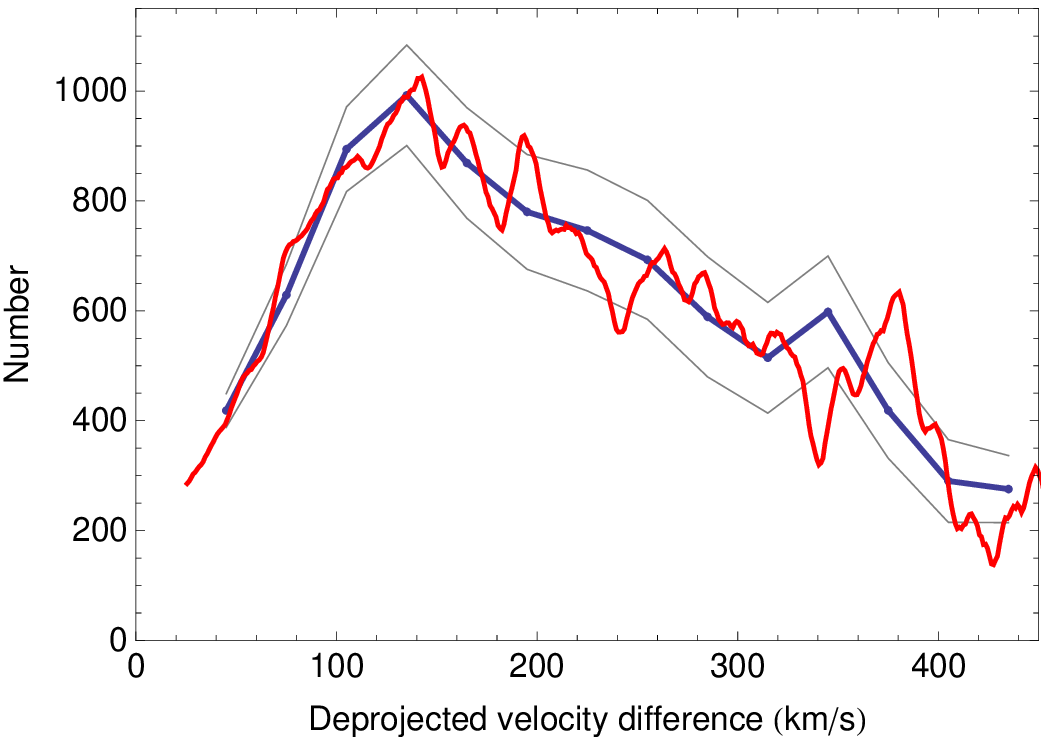} &
\includegraphics[width=7.5cm]{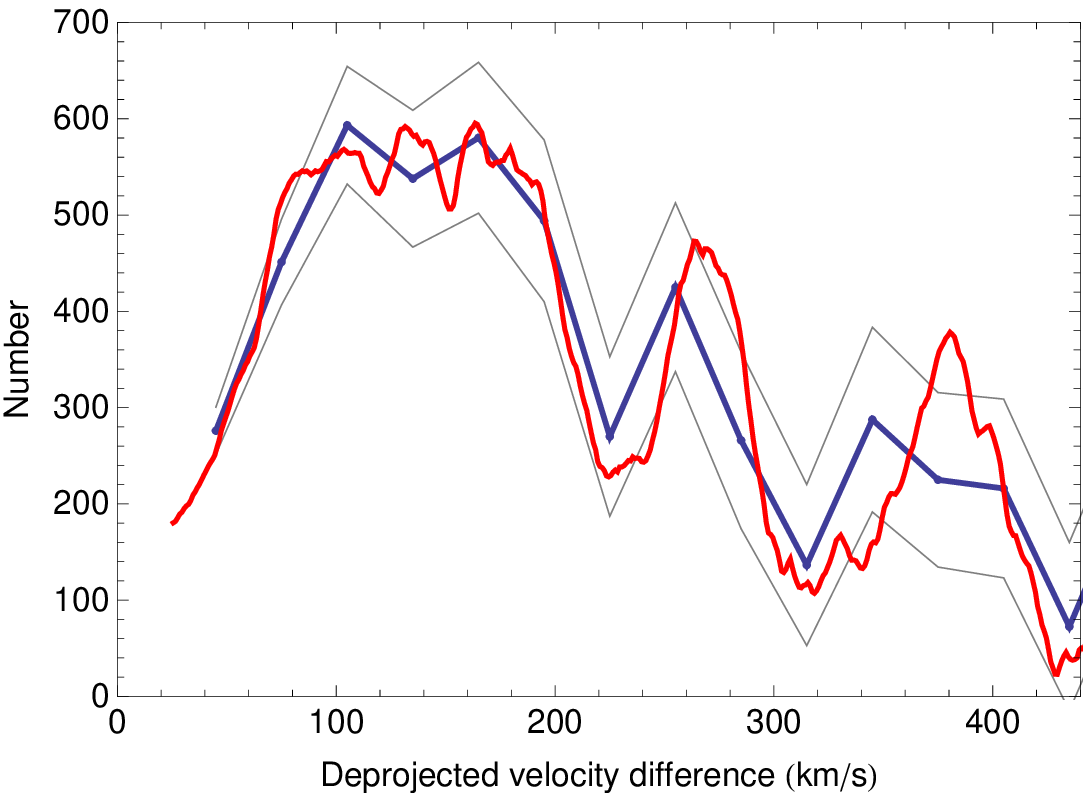}
\end{tabular}
\caption{{\small Deprojected PDF of the 3D true intervelocities between members of galaxy pairs in the IGP Catalog (all $\rho >2.5$ values). The blue curves are obtained by the deprojection method using two-bin differences  (bin of $30$ km/s). The red curves are the result of deprojection by moving bins (see text and Ref.~\cite{Nottale2018b}). The $\pm 1\sigma$ uncertainty curves (gray lines) are obtained from simulation of the deprojection process. Both methods agree within error bars. {\it Left figure:}  subsample of 11259 pairs with errors $<70$ km/s.  The projected distribution of these pairs is given in Fig.~\ref{Vproj}. {\it Right figure:} subsample of 6026 pairs having accurate velocities with errors $<20$ km/s. Both subsamples confirm the existence of a large probability peak around $\approx 150$ km/s, in agreement with the PDFs of fairly (Fig.~\ref{Vdeproj6}) and highly isolated pair (Fig.~\ref{Vdeproj5}) intervelocities.}}
\label{Vdeproj1}
\end{center}
\end{figure}

 Finally, we perform the deprojection of the whole catalog, containing all pairs having an isolation parameter $\rho>2.5$. We just exclude the 1855 pairs with inaccurate intervelocities $\delta V>70$ km/s.
We give in left Fig.~\ref{Vdeproj1} the result of this deprojection for the 11259 remaining pairs and in right Fig.~\ref{Vdeproj1} the deprojection of the subsample of 6026 pairs having accurate intervelocities ($\delta V<20$ km/s).

The existence of a main probability peak of orbital velocities around $150$ km/s, detected in fairly and highly isolated pairs, is confirmed for the whole catalog. However this peak is somewhat wider for all $\rho$'s than for the more isolated pairs $\rho>5$ and $\rho>10$. This result is not unexpected, since the highly isolated pairs can be considered as true 2-body Keplerian systems. The fairly isolated pairs can be slightly perturbed by a third body, while the perturbation may be larger for $2.5<\rho<5$. The unperturbed pairs exhibit a clear energy structure at $v_0^2$, with $v_0\approx 150$ km/s and ``line width'' $\approx \pm 50$ km/s. Whatever be the origin of this structure, one expects it to be subjected to a broadening effect in the perturbed pairs which are no more strictly 2-body systems.

Moreover,  possible secondary peaks are suspected in the highly isolated subsample, in particular around $\approx 360$ km/s. However this peak is not as stable as the $150$ km/s one, since it is absent in the fairly isolated subsample and marginally present in the whole sample. The reason for these fluctuations may be a combination of the effect of perturbations of the pair by other bodies, of radial velocity errors and of the fact that the projected probability density has a smaller value for higher velocities and is therefore subjected to higher relative uncertainties.

 \subsection{Deprojection of pair intervelocities in the UGC pair catalog}
Finally, the results obtained in the IGP catalog  are confirmed with  the UGC Galaxy Pair Catalog (UGCP). The deprojection of the projected intervelocity PDF for these pairs with highly accurate velocities, yields once again a clear main peak at $150$ km/s and a possible secondary peak at $\approx 350$ km/s.

\begin{figure}[!ht]
\begin{center}
\includegraphics[width=12cm]{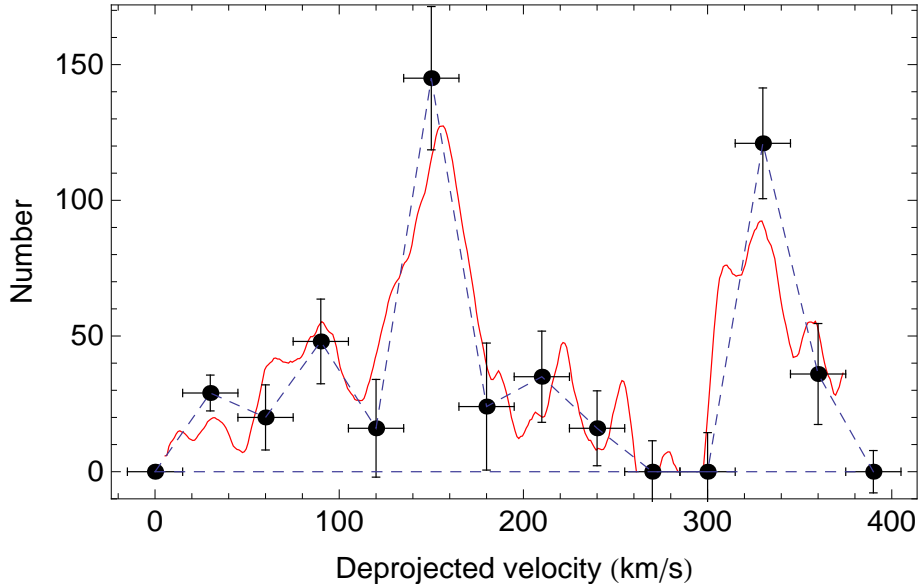}
\caption{{\small  Deprojected PDF of the 3D true intervelocities between members of galaxy pairs in the UGC pair Catalog, for the unbiased subsample  with blue major axes $<1.2$ arcmin. The red continuous curve is obtained by the deprojection method using a moving bin. The dashed black curve is the result of deprojection by constant bins of width $30$ km/s.  Both methods yield a main peak at $\approx 150$ km/s and a possible secondary peak at $\approx 350$ km/s, in agreement with the results from the IGP catalog.}}
\label{NilsonV}
\end{center}
\end{figure}

\section{Deprojection of pair interdistances}
\label{sec3}

\subsection{Statistical deprojection}
\label{sec3.1}
We have given in Ref.~\cite{Nottale2018b} the normalized probability density of $r_p$ values projected from a given $r$ value:
\beq
p(r_p)=\frac{dP(r_p)}{d r_p}=\frac{r_p}{r \sqrt{r^2-r_p^2}}
\eeq
(note the correction of a misprint in Eq. 7 of Ref.~\cite{Nottale2018b} in which the $r$ in the denominator, which allows the normalization, was lacking).
Therefore the probability density of projected distances $P_{r_p}(r_p)$ is given from that of true 3D distances $P_r(r)$ by the integral:
\beq
P_{r_p}(r_p)=\int_{r_p}^{\infty} \frac{P_r(r) \; r_p \; dr}{r \sqrt{r^2-r_p^2}}.
\eeq
Since it does not seem possible to invert this expression in the general case, we have constructed an algorithm to perform this inversion numerically \cite{Nottale2018b}. The deprojected distribution is recovered from the matrix product:
\beq
N_j^r= P_{ji}^{-1} N_i^p,
\eeq
where the column vector $N_i^p$ gives the projected number of pairs lying in the bin of rank $i$, the column vector $N_j^r$ gives the deprojected number of pairs lying in the bin of rank $j$ and the matrix  $P_{ji}$ is the transpose of:
\beq
\pi_{ij}=\sqrt{1-\l(\frac{i-1}{j}\r)^2}-\sqrt{1-\l(\frac{i}{j}\r)^2}
\eeq
for $i \leq j$.

\begin{figure}[!ht]
\begin{center}
\includegraphics[width=12cm]{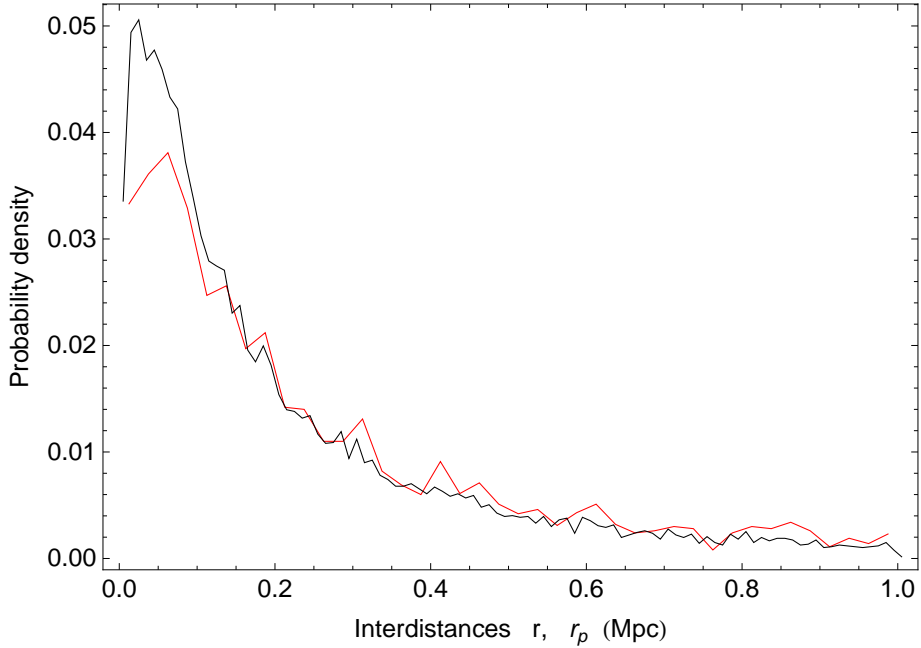}
\caption{{\small  Probability density distributions of the interdistances between members of galaxy pairs in the IGP Catalog. Black curve: observed (projected) PDF. Red curve: deprojected PDF, bias corrected (see text about the deprojection method). Both curves are normalized to an effective bin of $0.01$ Mpc.}}
\label{PDFsder}
\end{center}
\end{figure}

We give in Fig.~\ref{PDFsder} the observed distribution of the projected interdistances for the full IGP Catalog of 13114 pairs, obtained with a bin width of $0.01$ Mpc (100 bins between 0 and 1 Mpc). We have deprojected this PDF using the above matrix method applied to 40 bins of width $0.025$ Mpc. 

There is a small bias in this method since the vectors giving the projected and deprojected numbers are defined in the same range $0-1$ Mpc, while some projected pairs with $r_p<1$ Mpc may actually come from true interdistances $r>1$ Mpc. This artificially slightly increases the deprojected PDF for $r> \approx 0.8$ Mpc. We have corrected this bias by adding 4 bins beyond 1 Mpc in the projected PDF extrapolated from its power law fit (see hereafter). This results in the expected continuously decreasing distribution shown in Fig.~\ref{PDFsder}, where the bias has disappeared for $r<1$ Mpc.

The respective projected and unprojected probability densities are found to be remarkably identical, except for small interdistances $r$ and $r_p< \approx 0.1$ Mpc.

\subsection{Relation between projected and 3D interdistance PDFs}
\label{relation}
This similitude between the projected and deprojected PDFs can be easily understood from the fact that the true 3D PDF is very well fitted by a power law (except at very small scales), as can be seen in Figs.~\ref{PDFsder}, \ref{DeprojHubble}, and \ref{DeprojKing}. 

\begin{figure}[!ht]
\begin{center}
\includegraphics[width=12cm]{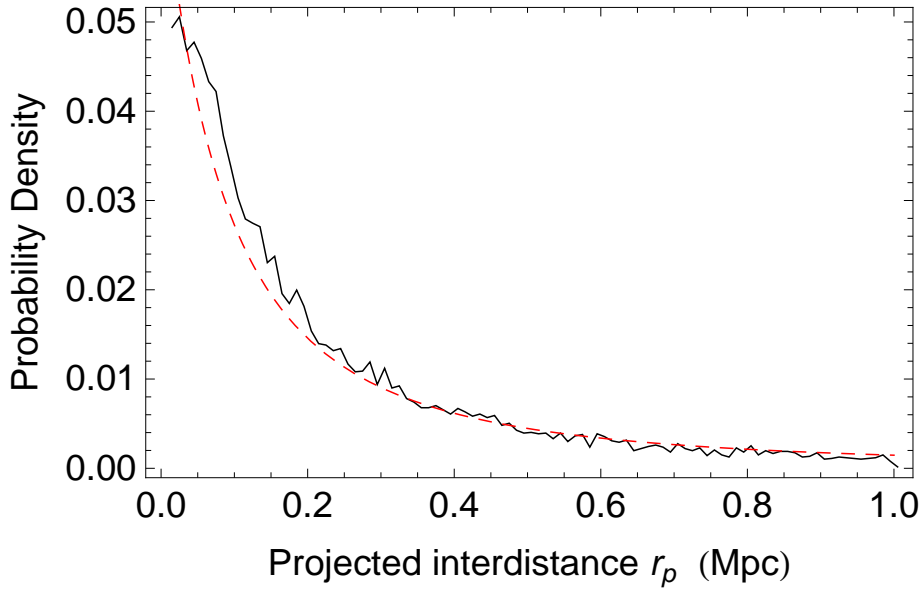}
\caption{{\small  Fit of the observed PDF of the projected interdistances between pair members of the IGPC by a one parameter Hubble profile $\propto (1+r_p/a_p)^{-2}$ (see Sec.~\ref{theor} for a justification of this law). The PDF is obtained from a normalized histogram with a bin of $0.01$ Mpc. The fitted parameter is $a_p=(0.17 \pm0.01)$ Mpc. }}
\label{ProjHubble}
\end{center}
\end{figure}

\begin{figure}[!ht]
\begin{center}
\includegraphics[width=12cm]{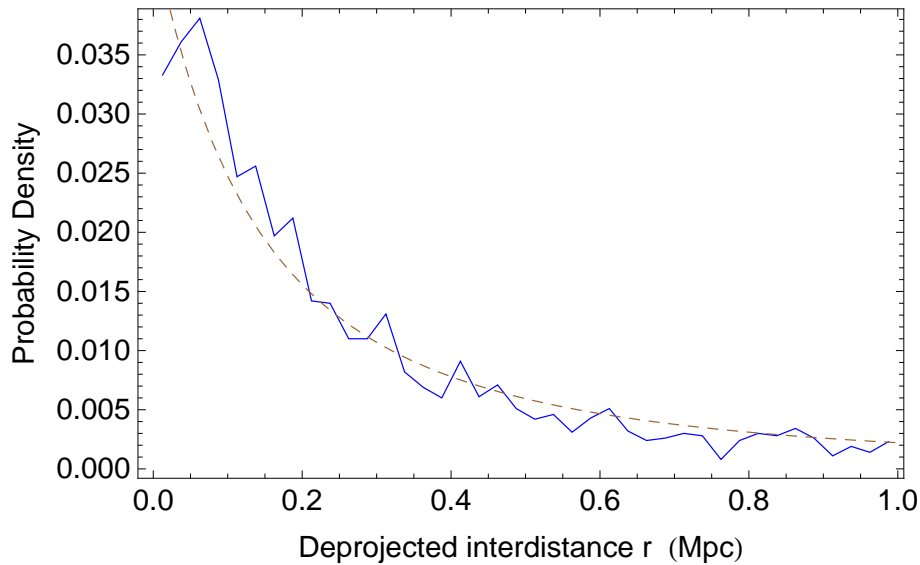}
\caption{{\small  Fit of the PDF of the deprojected interdistances between pair members of the IGPC by a one parameter Hubble profile $\propto (1+r/a)^{-2}$ (see Sec.~\ref{theor}). The statistical deprojection has been performed from a $r_p$ histogram with 40 intervals (bin of $0.025$ Mpc), but the normalized PDF is given with an effective bin of $0.01$ Mpc for comparison with the projected PDF. The fitted parameter is larger than for projected data (see Fig.~\ref{ProjHubble}), namely, $a=(0.28 \pm 0.02)$ Mpc. }}
\label{DeprojHubble} 
\end{center}
\end{figure}
Indeed, assuming a pure power law for the true interdistance between the pair members (strictly valid at large scales),
\beq
P(r) \propto \; r^{-2g},
\eeq
the projected interdistances PDF is given by  \cite{Nottale2018b}:
\beq
P_p(r_p) \propto \int_{r_p}^\infty \frac{r_p \; r^{-2g}}{r(r^2-r_p^2)^{1/2}} \: dr=\frac{\sqrt{\pi }\;  \Gamma \left(g+\frac{1}{2}\right)}{2 \; \Gamma (g+1)} \;  r_p^{-2g},
\eeq
where the function $\Gamma$ is the continuous generalization of the factorial function. Therefore, this theoretical argument confirms that if the deprojected PDF is a power law, the projected PDF is also given by a power law (both at large scales), with the same exponent (and reciprocally).

\begin{figure}[!ht]
\begin{center}
\includegraphics[width=12cm]{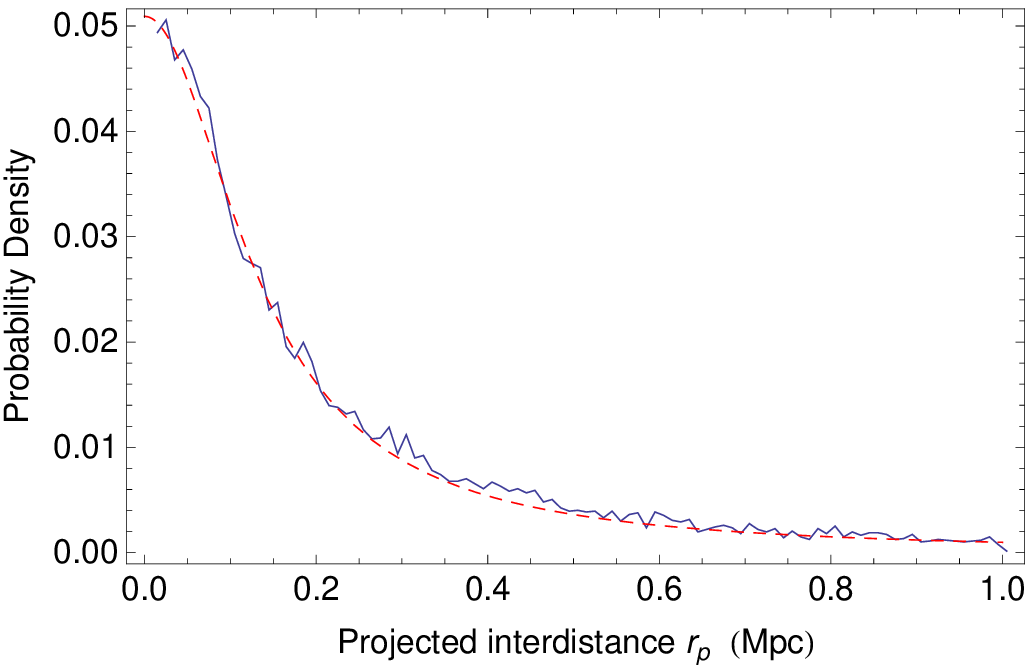}
\caption{{\small  Fit of the PDF of the projected interdistances between pair members of the IGPC by a two-parameter King profile $\propto ( 1+r_p^2 / r_{cp}^2 )^{-g_p}$.  The fitted parameters are $r_{cp}=0.133 \pm 0.004$ and $g_p=0.97 \pm 0.03$, which is compatible with $g_p=1$, the exponent expected from a theoretical argument involving a Laplace transform (see Sec.~\ref{theor}).}}
\label{ProjKing}
\end{center}
\end{figure}

\begin{figure}[!ht]
\begin{center}
\includegraphics[width=12cm]{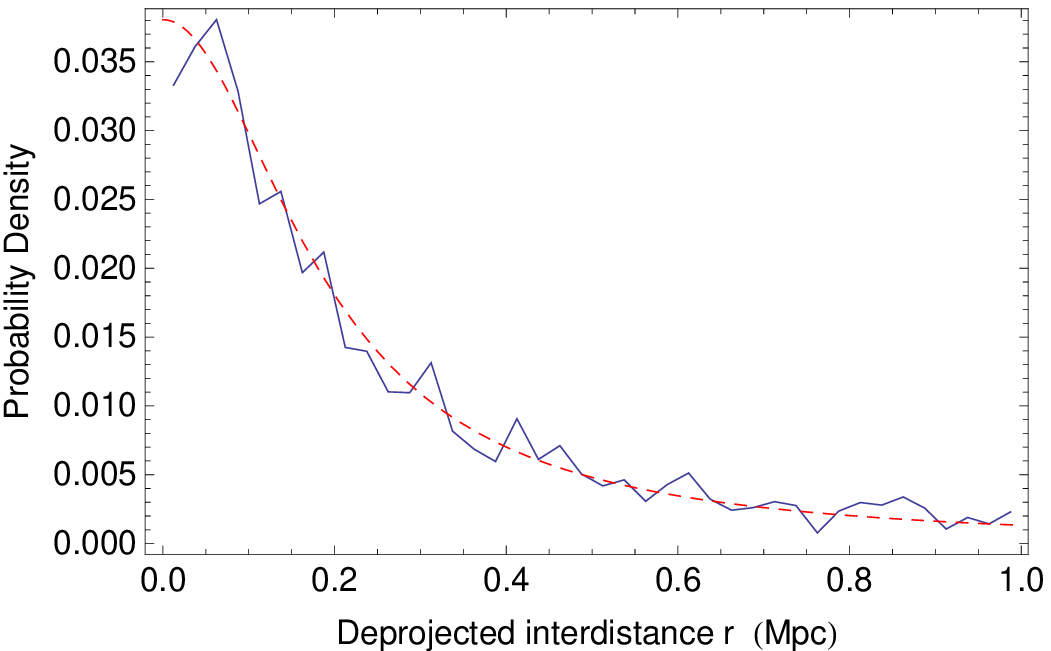}
\caption{{\small  Fit of the PDF of the deprojected interdistances between pair members of the IGPC by a two-parameter King profile $\propto ( 1+r^2 / r_c^2 )^{-g}$ for $r_c=0.19$ and $g=1$ (see  a justification of this law based on a Laplace transform in Sec.~\ref{theor}).}}
\label{DeprojKing}
\end{center}
\end{figure}

This is supported by the observed distributions: indeed, we can see in Fig.~\ref{PDFsder} that the projected and deprojected PDFs are indistinguishable for distances $r$ and $r_p>\approx 0.15$ Mpc: moreover the fit of the two distributions yields a similar exponent $g \approx 1$ (see Figs.~\ref{ProjHubble}, \ref{DeprojHubble}, \ref{ProjKing}, \ref{DeprojKing}).

Note that the IGP Catalog is limited to projected interdistances $r_p \leq 1$ Mpc while some pairs with $r>1$ Mpc may be projected to $r_p \leq 1$ Mpc. This explains why the upper limit of the above integral is taken to be infinity. We have corrected this bias in the previous Sec.~\ref{sec3.1}.

However, a power law is no longer physical at small scales due to its divergence at $r \to 0$ . We shall  therefore consider two improved models which remain a power law at large scales but include a cut-off at small scales. The first is a law long ago proposed by Hubble \cite{Hubble1930} for luminosity profiles of elliptical galaxies:
\beq
P(r)=\l(1+\frac{r}{a}\r)^{-2}.
\eeq

The second law considered including a core radius is a King profile \cite{King1966}:
\beq
P(r)= P_0 \l(1+\frac{r^2}{r_c^2}\r)^{-g},
\eeq
where $P_0$ is a normalization constant, which is given by:
\beq
P_0={_2F_1}\left(\frac{1}{2},g;\frac{3}{2};-\frac{1}{r_c^2}\right)^{-1},
\eeq
where $_2F_1$ is the hypergeometric function. It is well approximated in the relevant range of the variables (to better than $\approx 5 \%$) by $P_0 \approx  1+0.55 g/r_c$.

This PDF of true interdistances can be integrated to yield the PDF of projected interdistances:
\beq
P_{r_p}(r_p)=\frac{ \sqrt{\pi } \:  \Gamma \left(g+\frac{1}{2}\right)}
   {2 \;_2F_1\left(\frac{1}{2},g;\frac{3}{2};-\frac{1}{{r_c}^2}\right)}
   \;  \left(\frac{{r_p}}{{r_c}}\right)^{-2 g}
     {_2\tilde{F}_1\left(g,g+\frac{1}{2};g+1;-\frac{{r_c}^2}{{r_p}^2}\right)}.
\eeq
It is remarkable that we recover the basic power law $({r_p}/{r_c})^{-2 g}$, now corrected by an hypergeometric function. One can show that this exact solution is well approximated by another King profile with the same exponent $g=g_p$, but with a different core radius $r_c$  (except when $r \to 0$ where the King profile has a zero slope).

This result is supported by the observed, projected and deprojected distributions (Figs.~\ref{ProjKing} and \ref{DeprojKing}). A least-square fit yields $r_{cp}=0.13$. and $g_p=0.97 \pm 0.03$, which lies within $1 \sigma$ of $g_p=1$.  The deprojected distribution is well represented by a King profile with the same fixed exponent $g=1$, but with a core radius $r_c=0.19$, see Fig.~\ref{DeprojKing} (its least square fit yields more precisely $g = 0.87\pm 0.08$ and $r_c = 0.17 \pm 0.02$, which are compatible with these values).

\subsection{Theoretical expectation for the interdistance PDF}
\label{theor}

This asymptotically power law shape of the interdistance PDF can be derived from a simple (partly theoretical) argument based on the observed pair masses (derived from their luminosities) and intervelocity PDFs. 

\begin{figure}[!ht]
\begin{center}
\includegraphics[width=12cm]{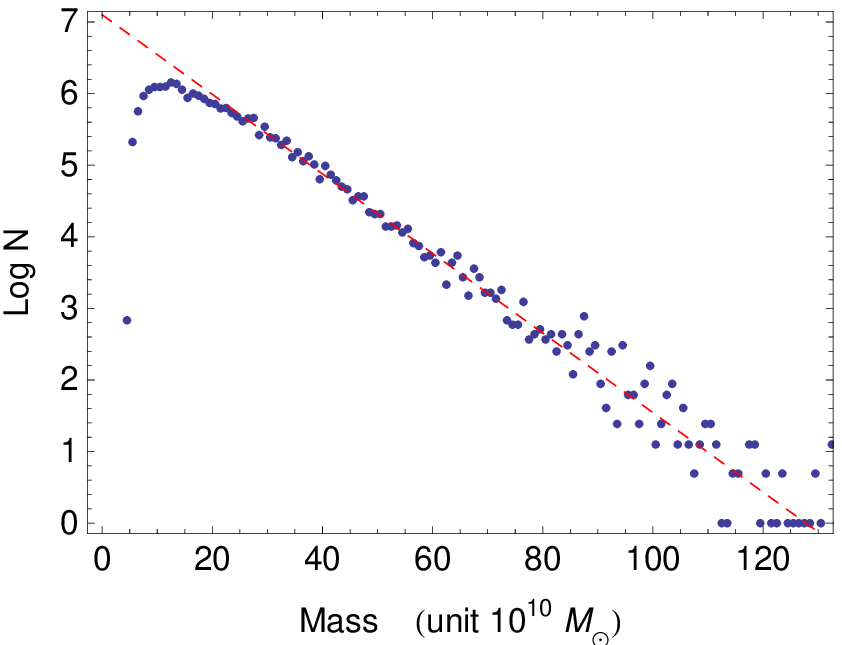}
\caption{{\small Fit by an exponential law of the PDF of pair masses (derived from total pair luminosities and a mean $M/L$ ratio of 30, as found from the mass deprojection performed in Sec.~\ref{mass}).}}

\label{PdeM}
\end{center}
\end{figure}

From Kepler third law $G M = r\; v^2$, one expects $r \propto M/v^2$. The luminosity PDF of pairs in the IGPC is found to be well fitted by a pure exponential law $P_L(L) \propto \exp(-L/L_0)$ except for small luminosities. Such a law is not incompatible with a standard Schechter law \cite{Schechter1976} for the most luminous pairs, as can be seen in Fig.~1 of Ref.~\cite{Nottale2018a}.
The same behavior is true for masses assuming a mean constant $M/L$ ratio (see Fig.~\ref{PdeM}). We have seen that the intervelocity PDF is systematically dominated by a probability density peak lying around $150$ km/s. It  can therefore be approximated by a Gaussian law $P_v(v)\propto \exp[-\frac{1}{2}(v-v_0)^2/\sigma_v^2]$, with $v_0\approx 150$ km/s and $\sigma_v \approx 85$ km/s.

\begin{figure}[!ht]
\begin{center}
\includegraphics[width=12cm]{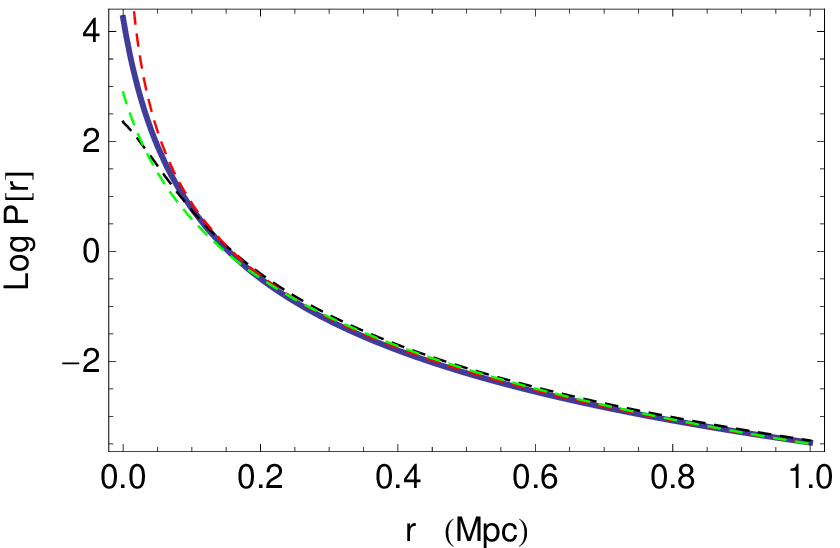}
\caption{{\small Analytical integration of the PDF of interdistances from an exponential mass PDF and a Gaussian intervelocity PDF, compared to various expectations. Blue curve: exact integral; red dashed curve: pure power law with exponent $-1.9$; black dashed curve: precise fit of deprojected data; green dashed curve: Hubble profile  $P(r)\propto (1+r/a)^{-2.1}$ with $a=0.05$. As expected, the analytical integral is no longer valid at very small distances $r<0.1$ Mpc.}}
\label{PdeR}
\end{center}
\end{figure} 

The PDF of $M/v^2$ can be derived from the individual PDFs of $M$ and $v$ provided they are independent. One finds, for an exponential PDF of mass:
\beq
P_r(r)= \int_0^\infty \exp\l(\frac{-r}{M_0} x\r)\; P_v(\sqrt{x}) \sqrt{x} \; dx,
\label{aaa}
\eeq
meaning that the interdistance PDF is the Laplace transform of the function $ \sqrt{x} \; P_v(\sqrt{x})$, where $x$ has the dimension of the square of a velocity. For a Gaussian velocity PDF $P_v$, one finds that this expression is fairly approximated by a function $\propto  x \exp(-x/w^2)$ (with $w\approx 200$ km/s for $v_0\approx 160$ km/s and $\sigma_v \approx 85$ km/s), the Laplace transform of which is well known to be a generalized power law
\beq
P(r)=\l(1+\frac{r}{a}\r)^{-2},
\eeq
which is similar to a Hubble luminosity profile.

\begin{figure}[!ht]
\begin{center}
\includegraphics[width=12cm]{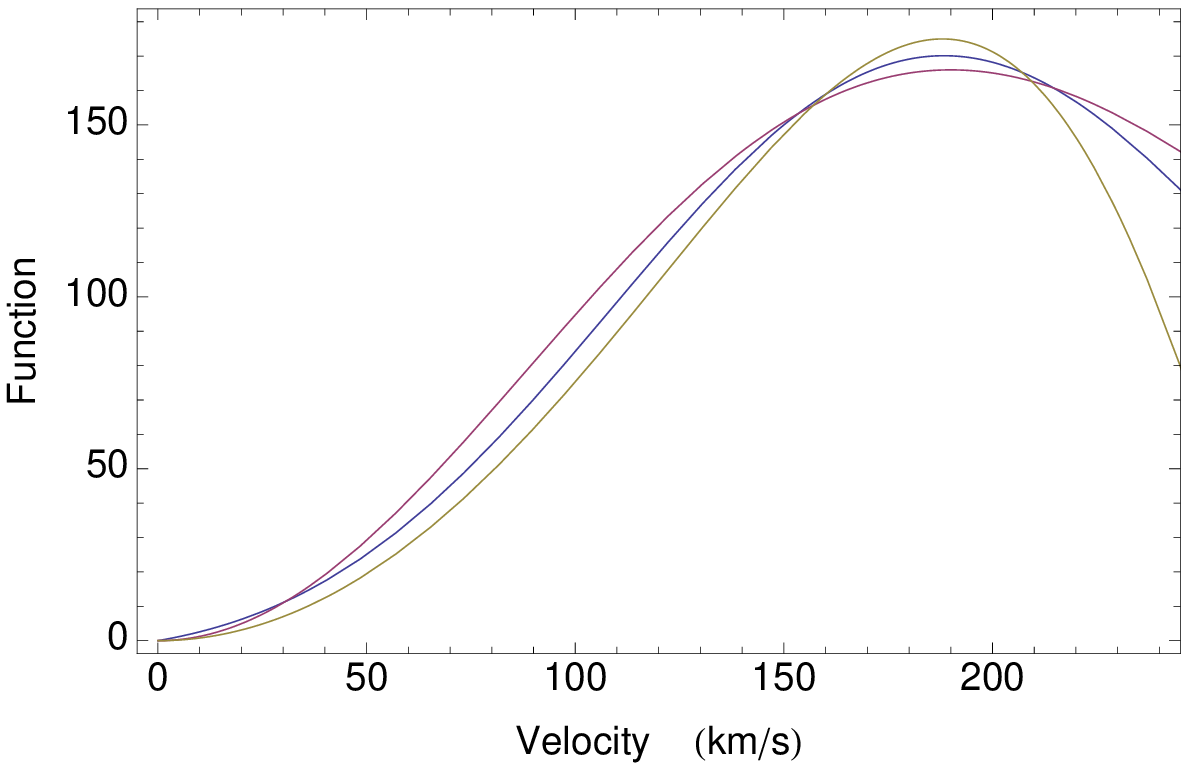}
\caption{{\small  Comparison between the function $ v \; P_v(v)$ (blue curve), which occurs in the derivation of the PDF $P_r(r)$ of pair members interdistances through a Laplace transform (see text), and  two close functions whose Laplace transforms yield respectively a Hubble interdistance profile (red curve) and a King interdistance profile (yellow curve).}}
\label{3functions}
\end{center}
\end{figure}

This is supported by a direct analytical integration of Eq.~\ref{aaa} whose result is shown in Fig.~\ref{PdeR}. Depending on the choice of the peak value and width of the Gaussian velocity PDF, one finds an excellent agreement of this analytical function with Hubble profiles having exponents in the range $1.9-2.2$ (except at small distances $r < 0.1$ Mpc). A similar result is obtained for more elaborated velocity PDFs made of a sum of normal distributions fitting the observed  deprojected PDF.


One may also support the obtention of a King profile by a similar argument. One can see in Fig.~\ref{3functions} that the three functions $v \exp[-\frac{1}{2}(v-v_0)^2/\sigma_v^2]$, $\propto v^2 \exp(-v^2/w^2)$ and $\propto \sin(v^2/u^2)$ are very close in the velocity range $[0-230]$ km/s, dominated by the main velocity peak around $150$ km/s (we have plotted them directly in terms of velocity $v$ instead of the variable which comes in the Laplace transform $x=v^2$). Now the Laplace transform of the function $\sin( \beta x)$ is just the King profile $\beta /(\beta^2+r^2)$  with exponent $g=1$.

Our results for the projected and deprojected PDFs of IGPC pairs, which have been shown to have the same asymptotic exponents, support this expectation of an exponent $2g \approx -2$ (see Figs.~\ref{ProjHubble} and \ref{DeprojHubble}). The fits by Hubble profiles (which are one-parameter functions) are, as expected, of lower quality than those by the two-parameter King profiles, but they remain fair knowing that they derive from a simple theoretical argument.

\section{Pair mass and M/L ratio}
\label{mass}
\begin{figure}[!ht]
\begin{center}
\includegraphics[width=12cm]{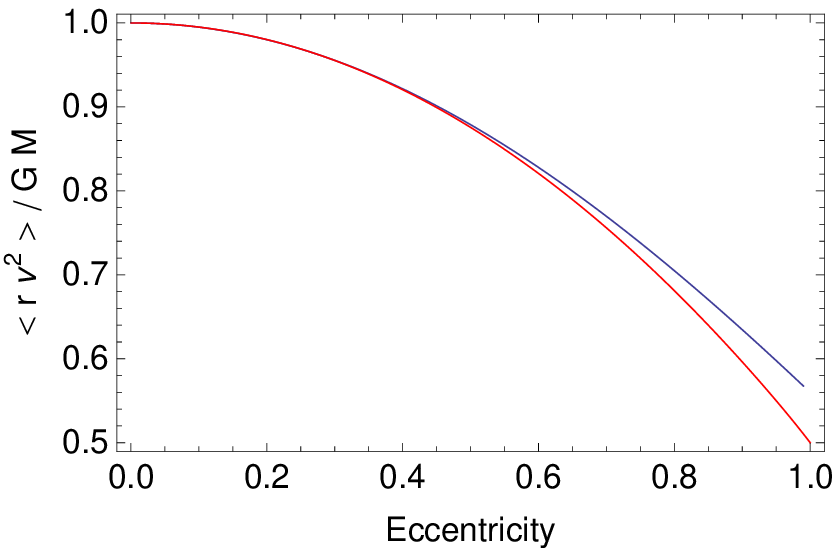}
\caption{{\small   Effect of eccentricity on the mass deprojection from Kepler's third law (see text).} }
\label{eccentricity}
\end{center}
\end{figure}

The mass $M$ of a galaxy pair is given from Kepler's third law by $ G M= a V^2$, where $a$ is the orbital semi-major axis of one body around the other and $V = 2 \pi a/T$, $T$ being the orbital period. In the circular case, $r=a$ = cst and $v=V$ = cst, so that the mass PDF can be derived from a deprojection of the product of observables $r_p \, v_z^2$ into $r \, v^2$. We have discussed the non-circular case in \cite{Nottale2018b}: one can show that  that the effect of not too large eccentricities on the mass determination remains small (see Fig.~\ref{eccentricity}).

\begin{figure}[!ht]
\begin{center}
\includegraphics[width=12cm]{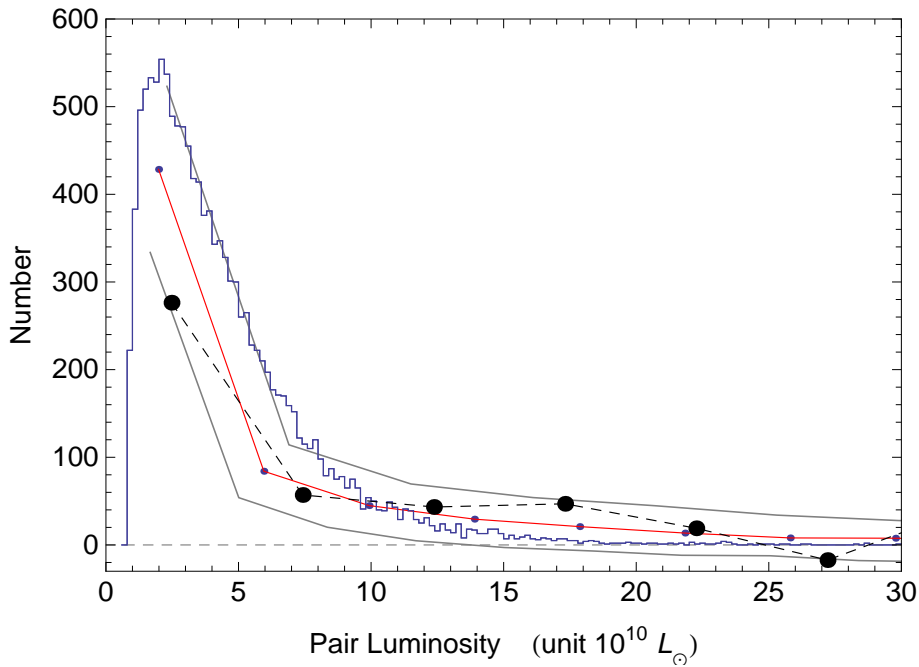}
\caption{{\small  Deprojection of the product  $r_p v_z^2$ (black points), compared with the total luminosity PDF of IGPC pairs (blue histogram).  The product of 3D variables $r v^2$ gives the total mass of the pair (in the circular orbit approximation). The resulting mass has been translated into luminosity through a constant $M/L$ ratio. The red line results from a numerical simulation which allows to settle the uncertainty of the deprojection (two black lines, $\pm 1$ sigma). The luminosity PDF falls  within $\approx$ one sigma of the deprojected points, allowing to derive a mass over luminosity ratio of $M/L=30 \pm 5$. } }
\label{Mdeproj1}
\end{center}
\end{figure}

We have given an original  deprojection method of $r_p \, v_z^2$ in Ref.~\cite{Nottale2018b} that uses a deprojection matrix, which has been confirmed by numerical simulations. Let us summarize it.
For a given value of $r$ and $v$ (one pair), the expected distribution of the product $\eta= r_p \, v_z^2$ is given by:
\beq
p_\eta(\eta)=\frac{ \Gamma(3/4)}{\Gamma(1/4) }  \, \sqrt{ \frac{\pi}{ \eta}}-\frac{\eta}{3} \; _2F_1\l(\frac{1}{2},\, \frac{3}{4},\, \frac{7}{4},\, \eta^2\r).
\label{Peta}
\eeq
For a population of pairs, having divided the $\eta$ range into $N$ bins, the projection matrix is written as
\beq
A_{ji}=\int_{(i-1)/(N+1-j)}^{i/(N+1-j)} p_\eta(\eta)\,  d \eta,
\eeq
where $j=1$ to $N$, $i=1$ to $N+1-j$ and $A_{ji}=0$ for the remaining coefficients. Finally the histogram of the deprojected product is given from the histogram of the projected product by the matrix relation  $P_{r v^2}=B \; P_{r_p v_z^2}$, where the deprojection matrix is:
\beq
B= {\rm Reverse[Inverse[Transpose[A]]}.
\eeq
We have applied this deprojection method to the IGP Catalog using a $20 \times 20$ matrix on the sub-sample of 11259 pairs having intervelocity errors $<70$ km/s.
 The result is given in Fig.~\ref{Mdeproj1} (black points and dashed lines), where the deprojected mass PDF has been converted into a luminosity PDF through the application of a constant $M/L$ ratio. The global shape obtained is compatible with the observed PDF of pair luminosities (a decreasing exponential) and they can therefore be fitted one to the other. This fit determines a $M/L$ ratio of $\mu=M/L=30 \pm 5$. 

The statistical agreement between the deprojected and observed PDFs is established  from a simulation achieved in the following way: from the deprojections of projected intervelocities and interdistances, we have constructed analytical models of the true velocities PDF (Gaussian or sum of Gaussian laws) and distances PDF (power law). Then we have performed several random projections of $v_z$ and $r_p$ from these PDFs, which has provided us with a sample of $Z_p=r_p \, v_z^2$ values. Finally we have applied our deprojection method to these values. The average and standard error calculated in this way (red and gray lines in Fig.~\ref{Mdeproj1}) includes within $\pm 1$ sigma for $M/L=30$ both the observed luminosity PDF and the deprojected $r_p \, v_z^2$ for the real pairs in the IGPC, which is a very satisfying agreement, despite the known difficulty of a reliable determination of pair masses \cite{Faber1979}. The error bar $\pm 5$ on the $M/L$ ratio is estimated from  varying the $M/L$ ratio and requiring that all deprojected points remain in agreement with the simulation (and therefore with the observed luminosity PDF).

\begin{figure}[!ht]
\begin{center}
\includegraphics[width=12cm]{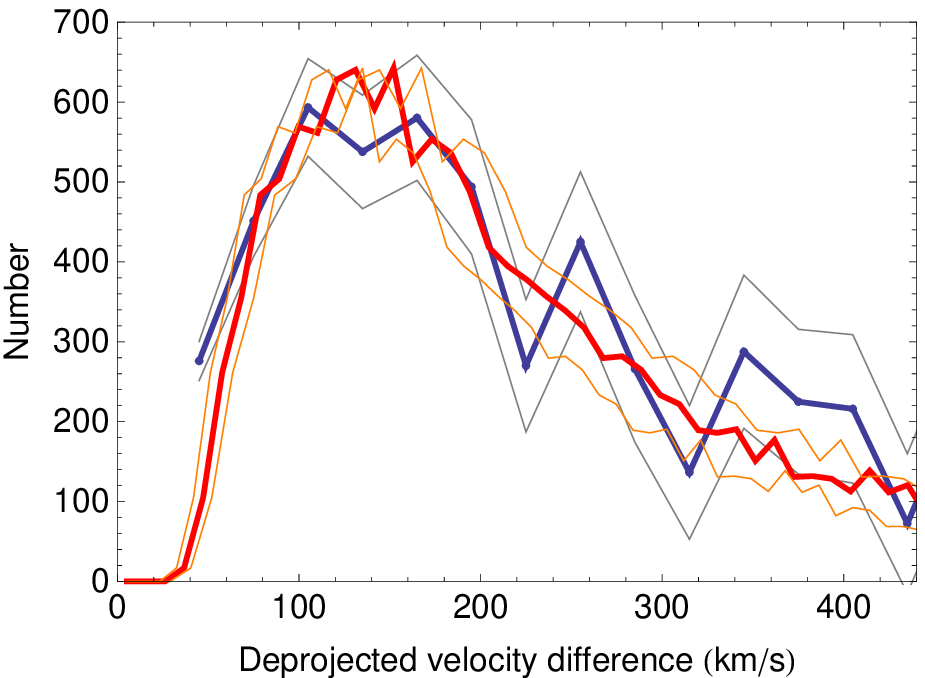}
\caption{{\small   Comparison between the PDF of the orbital velocity (6026 pairs with accurate radial velocities, blue curve with its error bars) and the PDF of $(G M / r_p)^{1/2}$, where the pair masses $M$ are obtained from their luminosity $L$ through a constant $M/L$ ratio, $M=\mu L$ (red curve). An excellent agreement between the two curves (everywhere within $\pm 1$ sigma) is obtained by fitting the $M/L$ ratio to the value $\mu=33 \pm 7$. The uncertainty has been estimated by varying the value of $\mu$ (orange curves, $\mu=26$ and $40$).} }
\label{vracGMsurr}
\end{center}
\end{figure}

In this first method, we have derived directly the mass from the relation $M= r v^2$ through the deprojection of $r_p \, v_z^2$ . A second method consists of writing this relation under the form $v= (M/r)^{1/2}= (\mu L/r)^{1/2}$ and to compare the observed deprojected velocity PDF with the deprojection of the PDF of $(L/r_p)^{1/2}$. The $M/L$ ratio $\mu$ can then be derived from this comparison, provided these two PDFs are found to be similar.  
This is indeed the case within uncertainties, as can be seen in Fig.~\ref{vracGMsurr}.

\begin{figure}[!ht]
\begin{center}
\includegraphics[width=12cm]{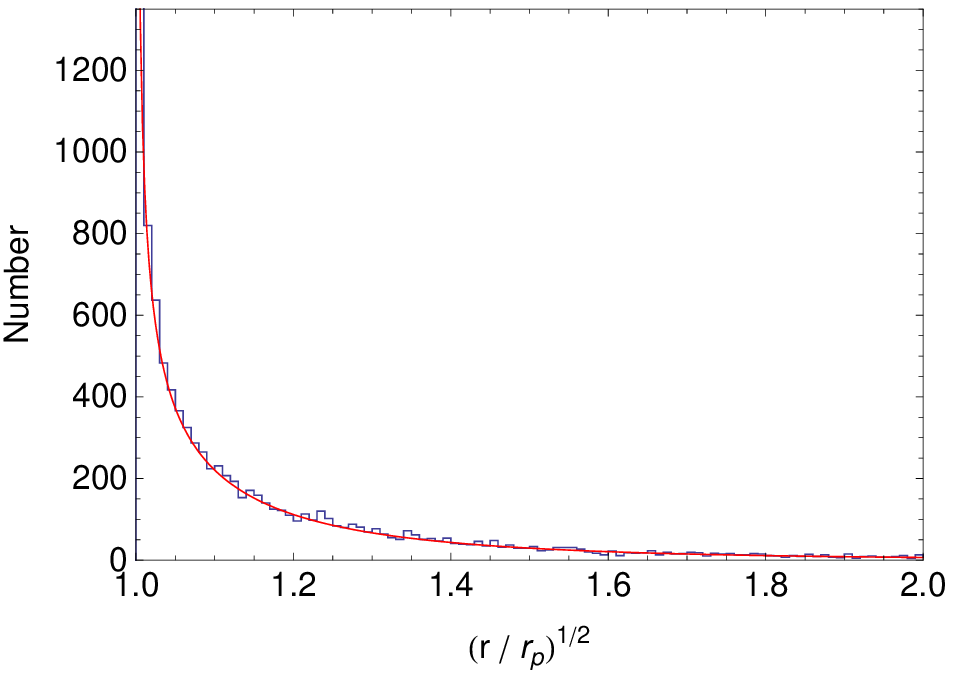}
\caption{{\small  Numerical simulation of the PDF of $(r/r_p)^{1/2}$ compared with its theoretical expression (Eq.~\ref{expr}). }}
\label{PDFrac1surrp}
\end{center}
\end{figure}

 The problem of finding the behavior of the PDF of $(M/r_p)^{1/2}$ is equivalent to finding the PDF of $r_p^{-1/2}$ provided there is no correlation between mass and interdistance. Actually the PDF of $r_p^{-1/2}$ shows a very narrow peak at $r_p=r$ (see Fig.~\ref{PDFrac1surrp}), thus implying a small projection effect. This is supported both by a numerical simulation and by the analytical expectation: one finds that the theoretically expected normalized PDF of this variable $y= r_p^{-1/2}$ for a given value of the 3D interdistance $r$ is:
 \beq
P_y(y) = \frac{2}{r \; y^3 \sqrt{r^2 y^4-1}}.
\label{expr}
\eeq
One can also show that the effect of eccentricity on $(GM/r)^{1/2}$ is weak and in the opposite way of its projection effect. This allows us to compare the velocity PDF directly with the PDF of $(G M / r_p)^{1/2}$ in Fig.~\ref{vracGMsurr}, since it is very close to the deprojected PDF of $(G M / r)^{1/2}$ (a fact that we have explicitly checked).

The $M/L$ ratio obtained by this second method, $\mu=33 \pm 7$ fairly confirms the value of the first method (direct deprojection of mass through $r_p \, v_z^2$),  $\mu=30 \pm 5$. 

\section{Discussion}
\label{discussion}

One of the main results of the present paper is the evidence of a systematic probability peak at $\approx150$ km/s in the PDF of pair galaxy orbital (deprojected) velocities. Now these isolated pairs achieve one of the simplest possible Keplerian systems, the two-body problem, at extragalactic scales of ($0.01-1$) Mpc. Expressed in terms of reduced mass, it becomes equivalent to a central mass problem with $m<M$.

It is therefore interesting to compare these systems to the archetype of Keplerian systems, i.e. to the numerous star-planet couples now found in exoplanet searches. It is already remarkable that, despite the $10^{11}-10^{12}$ factor between their interdistances (from $\approx$ AU to Mpc), the magnitude of their orbital velocities is quite similar, of the order of  $\approx 100$ km/s. This can be understood from the equivalence principle, according to which the inertial mass disappears from the equation of purely gravitational dynamics. If there are structures, we therefore expect them to occur in an universal way in velocity-space (energy structures become $E/m \sim v^2$ and momentum structures become $p/m=v$), while those in position-space are derived through the intervention of mass: one indeed recovers the above interdistance factor as being nothing else than the mass factor $10^{11}-10^{12}$ (for our sample) between galaxy masses and typical star masses ($\approx$ solar masses).

\begin{figure}[!ht]
\begin{center}
\includegraphics[width=12cm]{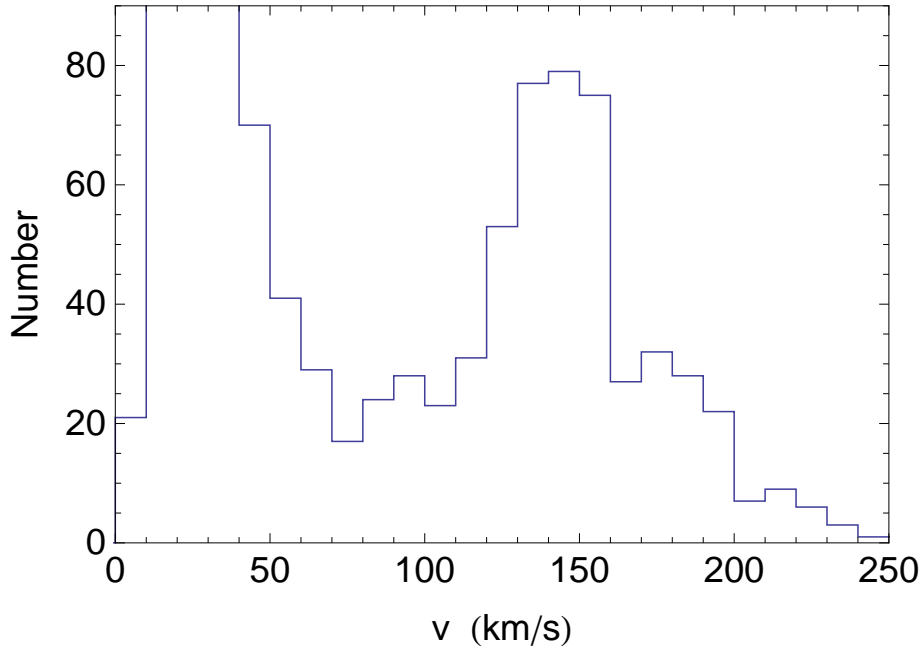}
\caption{{\small  Observed distribution of orbital velocities of $1183$ exoplanets having masses $>0.05$ 
Jupiter mass, extracted from the NASA Exoplanet Archive \cite{NASA2019} (June 2019 data containing 4104 confirmed planets). This PDF exhibits a large probability peak at a velocity of $\approx 150$ km/s, which is the same value as the main orbital velocity peak of galaxies in pairs. }}
\label{exoplanets}
\end{center}
\end{figure}

But there is better: not only the orbital velocity magnitude is the same, but exoplanets, as galaxy pairs, show a probability peak at just the same velocity value, i.e. around $150$ km/s. We have plotted in
Fig.~\ref{exoplanets} the distribution of orbital velocities of exoplanets (2019 data) having masses $>0.05$ Jupiter mass. This distribution shows two main peaks, one at velocities $\approx 20-50$ km/s, which are just typical of the inner solar system and the other one at $\approx 150$ km/s. It is remarkable that this value is just the velocity of the first exoplanet discovered around a solar-type star, 51 Peg \cite{Mayor1995} and that it has been predicted from a planetary formation model fitted to the inner solar system structure, before the actual discovery of exoplanets \cite[Chap.7]{Nottale1993}. Since then, a large number of exoplanets continued to contribute to this probability peak (see Fig.~\ref{exoplanets}), which can be shown to be robust against the various biases affecting exoplanet data.

This result points toward a possible universality of Keplerian structures in velocity space, whatever the interdistance or the mass scales \cite{Nottale1993,Nottale1996,Nottale2000,Nottale2011}.

As regards our determination of pair masses, we have found a ratio $M/L=30 \pm 5$, which confirms the standard value for galaxies and their halos. But here our result is far more accurate than the previous estimates of pair masses, thanks to our fit of the whole deprojected PDF, instead of using only averages (which revealed to be highly biased, due to the strong non-Gaussianity of the distribution). This corresponds to a dark matter contribution about 5 times the luminous matter (taking the standard mean value of $M/L=6$ for stellar luminous matter). This means that no non-baryonic contribution is needed at the scale of galaxy pairs. Such a result is likely to be mainly due to the presence of large interpenetrating halos around each pair member.

\section{Conclusion}
\label{conclusion}

This paper was mainly devoted to the statistical deprojection of intervelocities, interdistances and masses (through Kepler's third law) in two galaxy pair catalogs, the Isolated Galaxy Pair Catalog (IGPC) containing more than 13000 pairs, completed by the UGC pair catalog ($\approx 1000$ pairs).

The deprojected PDF of pair intervelocities has been found here to be systematically dominated by a probability peak at $\approx 150$ km/s. This result is obtained for both catalogs and all selected subsamples.  The fact that there exists a probability peak at the same orbital velocity for exoplanets suggests a possible universality of Keplerian structures whatever the spatial and mass scales.

The deprojected PDF of galaxy interdistances in pairs has been found to be fairly described by a power law of exponent $-2$ with a cut-off at small distances, well fitted by a King profile. This behaviour allows the projected PDF to be described by a very similar King law. Moreover, this power law can be analytically derived through a Laplace transform from the observed PDF of pair luminosities (and therefore of pair masses for a constant $M/L$ ratio), which we find to be accurately given by a decreasing exponential, and from the obtained PDF of deprojected (orbital) velocities.

We have obtained the deprojected mass PDF from velocities and distances through Kepler's third law by two different and complementary methods and we have found it to fairly agree with the shape of the observed luminosity PDF. This result has allowed us to calculate an accurate mass / luminosity ratio $M/L=30 \pm 5$ by fitting the two distributions with each other.

The present paper was focused on deprojection of observational data and analysis of the deprojected data on galaxy pairs. The physical implications of our results will be considered in a forthcoming work. We also intend to extend the statistical deprojection methods to other orbital elements, in particular to eccentricities and semi-major axes.

{\bf Acknowledgements} This research has made use of the NASA Exoplanet Archive, which is operated by the California Institute of Technology, under contract with the National Aeronautics and Space Administration under the Exoplanet Exploration Program.



\end{document}